\documentclass[twocolumn,showpacs,preprintnumbers,amsmath,amssymb]{revtex4}
\usepackage{graphicx}% Include figure files
\usepackage{dcolumn}% Align table columns on decimal point
\usepackage{bm}% bold math

\begin{document}

\title{Indication of a Co-Existing Phase of Quarks and Hadrons in Nucleus - Nucleus Collisions} 

\author{Bedangadas Mohanty, Jan-e Alam, Sourav Sarkar, Tapan K. Nayak and Basanta K. Nandi }

\medskip

\affiliation{Variable Energy Cyclotron Centre, Calcutta
   700064, India}

\date{\today}

\begin{abstract}

The variation of average transverse mass of identified
hadrons with charge multiplicity have been studied for AGS, SPS and 
RHIC energies. The observation of a plateau in the average transverse 
mass for multiplicities corresponding to SPS energies is attributed to 
the formation of a co-existence phase of quark gluon plasma and hadrons. 
A subsequent rise for RHIC energies may indicate a deconfined phase in the
initial state.
Several possibilities which can affect the average transverse mass are 
discussed. Constraints on the initial temperature and  
thermalization time have been put from the various experimental 
data available at SPS energies.

\end{abstract}

\maketitle

Results based on QCD (Quantum Chromodynamics) renormalization group approach 
predict that strongly interacting systems at high temperatures and/or densities
are composed of weakly interacting quarks and gluons~\cite{collins,kisslin} 
due to asymptotic freedom and the Debye screening of color charge.
Nucleus-Nucleus (A-A) collisions at very high energies may 
create situations conducive for the formation of a thermodynamic state 
where the properties of the system 
are governed by quarks and gluonic degrees of freedom. Such a state
is called  quark gluon plasma (QGP).  QCD  
lattice gauge theory suggest that the critical temperature ($T_c$) 
for such a transition is $\sim 170$ MeV~\cite{lattice}.
This has led to intense theoretical 
and experimental activities in this field of research~\cite{qm}. 

Experimental detection of the QGP in A-A collisions 
is a non-trivial task because of the small space-time volume of the 
system.  Among many signals of the QGP formation~\cite{probe}, one of the 
earliest is based on the relation of the thermodynamical variables, 
temperature and entropy to the average transverse momentum and multiplicity, 
respectively.
This was originally proposed by Van Hove in the context of proton-proton 
collisions~\cite{vanhove}. It was argued that a plateau
in the transverse momentum beyond a certain value of multiplicity will indicate
the onset of the formation of mixed phase of QGP and hadrons; analogous to
the plateau observed in the variation of temperature with entropy in a first 
order phase transition scenario. One hence looks for the
variation of average transverse momentum 
($\langle p_{T} \rangle$) or transverse mass ($\langle m_{T} \rangle$, 
$m_T=\sqrt{p_T^2+m^2}$) with respect to the total number of particles 
produced per unit rapidity ($dN/dY$) in A-A collisions at high energies.  
An increase in energy density
should increase  $\langle m_{T} \rangle$ till it reaches a 
critical density for phase transition to occur, where it should then 
show a plateau due to formation of mixed phase. Further increase of energy 
density should again increase the $ \langle m_{T} \rangle$ of 
produced particles. 

Several attempts have been made so far to look for such signals.
While the data from cosmic ray experiments~\cite{cosmic} are 
inconclusive, the data from A-A collisions in the laboratory for
fixed center of mass energies ($\sqrt{s}$) with different 
centralities (impact parameters)~\cite{wa98,rhic1,rhic2} 
have so far not shown this kind of behaviour.
This implies that a mere change in the centrality of the
collisions does not change the energy density of the system
formed after the collisions required to create any change 
of phase.  So it is imperative to study the 
variation of $\langle m_{T} \rangle$
with $dN_{ch}/dY$ for a broad range of beam energies for
fixed centrality.

In this letter the variation of $\langle m_{T} \rangle$
with $dN_{ch}/dY$ is examined for AGS, SPS and RHIC energies
spanning $\sqrt{s}$ from 2A GeV to 200A GeV.
We then carry out explicit theoretical calculations to understand the 
observed behaviour in terms of the properties of the matter produced in
the initial stages. 
\begin{figure}
\vskip -2.0cm
\begin{center}
\includegraphics[scale=0.4]{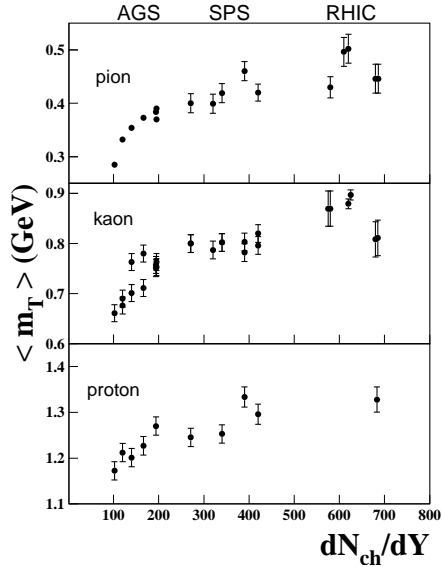}
\caption{ 
Variation of $\langle m_{T} \rangle$ with produced charged particles per
unit rapidity at mid rapidity for central collisions corresponding to
different $\sqrt{s}$ spanning from from AGS to RHIC. 
The error bars reflect both the systematic and statistical errors in
obtaining $T_{eff}$.
}
\label{fig1}
\end{center}
\end{figure}

In fig.~\ref{fig1} the variation of 
$\langle m_{T} \rangle$ with charge multiplicity is
depicted for pions, kaons and protons  
at AGS, SPS and  RHIC energies around mid-rapidity.  
The data shown here correspond to central events for
different colliding systems, (produced) particle types and 
center of mass energies (see Table~\ref{table1}).
%The sources of the experimental data are shown in the last column 
%of Table~\ref{table1}.
The experimental data on transverse mass spectra can be parametrized  as, 
\begin{equation}
\frac{dN}{m_{T}~dm_T}\,\sim\,C\exp\left(-\frac{m_{T}}{T_{eff}}\right)~,
\label{eq1}
\end{equation}
where the inverse slope parameter $T_{eff}$ is the effective temperature
(effective because it includes the contribution from both the thermal 
and collective motion in the transverse direction). 
The variation of $T_{eff}$ with $\sqrt{s}$ for kaons has recently been   
shown in Ref~\cite{marek}.
The average transverse mass of the particles obtained
from Eq.~\ref{eq1} is:
\begin{equation}
\langle m_{T}\rangle ~=~ T_{eff}~+~m~+~\frac{(T_{eff})^{2}}{m~+~T_{eff}}~.
\label{eq2}
\end{equation}

\begin{table}
\caption{The details of the experimental data used in the analysis
along with the references.
\label{table1}}
\begin{tabular}{lcccccr}
\tableline
$\sqrt{s}$ (GeV)&Type&centrality& $m_T$-$m_{\circ}$&Ref.\\
\tableline
2.3  (Au+Au)&$\pi^{+},K^{+}$      &$5\%$   & 0.1-1.0,0.05-0.7  &~\cite{ags}\\
2.3  (Au+Au)& p                   &$5\%$   & 0.0-1.0           &~\cite{ags2}\\
3.0  (Au+Au)&$\pi^{+},K^{\pm}$    &$5\%$   & 0.1-1.0,0.05-0.7  &~\cite{ags}\\
3.0  (Au+Au)& p                   &$5\%$   & 0.0-1.0           &~\cite{ags2}\\
3.6  (Au+Au)&$\pi^{+},K^{\pm}$    &$5\%$   & 0.1-0.9,0.05-0.7  &~\cite{ags}\\
3.6  (Au+Au)& p                   &$5\%$   & 0.0-1.0           &~\cite{ags2}\\
4.1  (Au+Au)&$\pi^{+},K^{\pm}$    &$5\%$   & 0.1-1.0,0.05-0.7  &~\cite{ags}\\
4.1  (Au+Au)& p                   &$5\%$   & 0.0-1.0           &~\cite{ags2}\\
4.7  (Au+Au)&$\pi^{+},K^{\pm}$    &$5\%$   & 0.1-0.8,0.05-0.7  &~\cite{ags}\\
4.86 (Au+Au)&$\pi^{\pm},K^{\pm}$  &$10\%$  & 0.1-1.5,0.05-0.8  &~\cite{ags1}\\
4.86 (Au+Au)& p                   &$7\%$   & 0.05-1.0          &~\cite{ags1}\\
8.76 (Pb+Pb)&$\pi^{-},K^{\pm}$    &$7.2\%$ & 0.2-0.7,0.05-0.9  &~\cite{sps}\\
8.76 (Pb+Pb)& p                   &$7.2\%$ & 0.0 - 1.0         &~\cite{sps}\\
12.3 (Pb+Pb)&$\pi^{-},K^{\pm}$    &$7.2\%$ & 0.2-0.7,0.05-0.9  &~\cite{sps}\\
12.3 (Pb+Pb)& p                   &$7.2\%$ & 0.0 - 1.0         &~\cite{sps}\\
17.3 (Pb+Pb)&$\pi^{-},K^{\pm}$    &$5\%$   & 0.2-0.7,0.05-0.9  &~\cite{sps}\\
17.3 (Pb+Pb)& p                   &$5\%$   & 0.0 - 1.0         &~\cite{sps}\\
17.3 (Pb+Pb)&$\pi^{-},K^{-}$      &10$\%$  & 0.1-1.2,0.05-1.0  &~\cite{sps1}\\
17.3 (Pb+Pb)&$\pi^{-},K^{\pm}$    &$3.7\%$ & 0.28-1.2,0.05-0.84 &~\cite{sps2}\\
130  (Au+Au)&$\pi^{\pm},K^{\pm}$  &$5\%$   & 0.1-1.0,0.1-1.2   &~\cite{rhic}\\
130  (Au+Au)& p                   &$5\%$   & 0.05-2.0          &~\cite{rhic}\\
130  (Au+Au)&$\pi^{-},K^{\pm}$    &$5\%,6\%$ & 0.02-0.6,0.05-1.2 &~\cite{rhic1,rhic2}\\
200  (Au+Au)&$\pi^{\pm},K^{\pm}$  &$10\%$  & 0.28-1.4,0.1-1.2 &~\cite{rhic3}\\
\tableline
\end{tabular}
\end{table}

From the results shown in fig~\ref{fig1}, one observes an 
increase in $\langle m_T\rangle$ with $dN_{ch}/dy$ for AGS
energies followed by a plateau 
for charge multiplicities corresponding to 
SPS energies for all the particle types, pions, kaons and protons.
This may hint at the possible co-existence of the quark and hadron phases.
For charge multiplicities corresponding to RHIC energies, the 
$\langle m_{T}\rangle$ shows an increasing trend
indicating the possibility of a pure QGP formation.

It is essential now to investigate whether the above 
experimental variation of 
$\langle m_{T}\rangle$ can arise from various physical effects other than
due to QGP formation. We will address this issue by trying to answer the 
following questions :
(a) How does (3+1) dimensional hydrodynamical evolution of the
system, with and without QCD phase transition, formed in heavy ion 
collisions affect the $\langle m_{T}\rangle$ at freeze-out ?
(b) Does the analysis of experimental data on hadrons, photons
and dileptons indicate similar values of the initial temperature ($T_i$)? 
How does the value of
$T_i$ compare with the  $T_c$ predicted by lattice QCD~\cite{lattice}?
(c) How does the $\langle m_{T}\rangle$ get affected
by the gain in transverse momentum by hadrons through their 
successive collisions (Cronin effect~\cite{cronin}) 
with the particles in the system ?
(d) How does the ${\langle m_{T}\rangle}$ of produced
particles vary with $\sqrt{s}$ in a standard event generator based
on the principle that nucleus-nucleus collision is a superposition of
nucleon-nucleon collisions ?

(a) To understand the variation of $\langle m_{T}\rangle$ 
with $dN_{ch}/dY$ for a system undergoing a 
(3+1) dimensional expansion we solve the hydrodynamical
equations for various initial energy densities ($\epsilon_i$)~\cite{hvg}.
The initial radial velocity is taken as zero.   
The momentum distribution of particles originating from a system 
undergoing transverse expansion~\cite{hvg} with boost invariance along
the longitudinal direction~\cite{jdb}, contains the effect of thermal
motion as well as transverse flow. Hence it is imperative to examine
how the value of the inverse slope or average transverse mass is affected
by transverse motion. To achieve this we proceed as follows. We evaluate 
the $m_T$ spectra for various hadrons at the freeze-out point as a 
function of $\epsilon_i$  for two different kind of space time 
evolution scenarios. 

\begin{figure}
\vskip -2.0cm
\begin{center}
\includegraphics[scale=0.4]{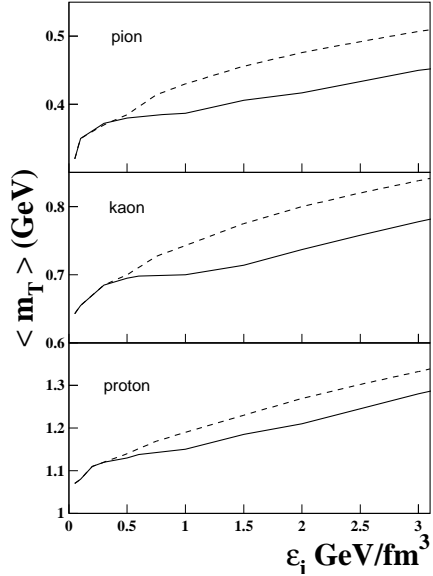}
\caption{ 
Variation of $\langle m_{T} \rangle$ with initial energy 
density ($\epsilon_{i}$) for a QGP scenario (solid line) and hadronic scenario 
(dashed line) obtained from (3+1) dimensional hydrodynamics.
}
\label{fig2}
\end{center}
\end{figure}

Firstly, we consider a hadronic gas at various values
of $\epsilon_{i}$ from $\sim 0.05 $ to 4 GeV/fm$^3$ without introducing a
QCD phase transition.
For all these initial energy densities the (3+1) 
dimensional hydrodynamic equations are solved to obtain the freeze-out
surface at a temperature of 120 MeV. This is used as an input
to evaluate the transverse mass spectra of hadrons. The hadronic equation
of state (EOS)~\cite{bm} contains hadronic degrees of freedom up to strange 
sector~\cite{pdg}. From the $m_T$ spectra of hadrons (pions, kaons 
and protons) corresponding to each $\epsilon_i$ we evaluate the
$\langle m_{T}\rangle$. This is plotted in Fig.~\ref{fig2} (dashed line).

In the second scenario, we introduce the QGP (composed of up, down,
strange quarks and gluons) at an energy density above $\epsilon_Q\sim 1.7$
GeV/fm$^3$ corresponding to critical temperature, $T_c=170$ MeV which converts
to the hadronic phase at an energy density, $\epsilon_H\sim 0.56$ GeV/fm$^3$
at the same temperature. The effective degeneracy ($g_{eff}$) at $T_c$ for 
the hadronic phase is $\sim 16$ obtained from the study of hadronic 
EOS in ~\cite{bm}.  
So in this scenario
we have a mixed phase within the energy range 
$0.56\,<\,\epsilon\,$(GeV/fm$^3$)$<\, 1.7$.
For $\epsilon>1.7$GeV/fm$^3$ a QGP bag model EOS has been used. 
The variation of 
$m_T$ as a function of $\epsilon_i$ for this scenario is shown 
in Fig.~\ref{fig2} (solid line). 
It is clear
that $\langle m_T\rangle$ varies at a much slower rate for 
the second scenario. 
Although a complete plateau structure is not observed,
it can be argued that the formation of a mixed phase slows down the rate 
of increase of $\langle m_{T} \rangle$ with multiplicity.
In fact, the slope of the curve in the case of pions for the first 
scenario is larger than the second
by a factor of 2.5 at $\epsilon_{i}$ = 1 GeV/$fm^{3}$.
The $\langle m_{T} \rangle$ 
for pions is about 400 MeV
for values of $\epsilon_i$ corresponding to mixed phase,
which is similar to those obtained experimentally.

(b) The lower bound on the $T_i$ realized at SPS energies
can be obtained from the measured $\langle m_{T}\rangle$ provided
we can estimate the effects of the flow.
The inverse slope of the $m_T$ spectra contains the effect 
of flow and can be parametrized as~\cite{uh},
\begin{equation}
T_{eff}=T\sqrt{\frac{1+\langle v_r\rangle}{1-\langle v_r\rangle}}
\label{blue}
\end{equation}
where $\langle v_r\rangle$ is the average velocity and $T=T_F$ is the
``true'' freeze-out temperature. 
We start by noting that $\langle m_{T}\rangle$
is actually the average energy, $\langle E\rangle$ of the particle 
at zero rapidity ($E=m_T\,coshy$) and in a thermal system the average
energy is proportional to temperature. Therefore, for a system which
has zero transverse velocity but non-zero longitudinal
velocity one must have
\begin{equation}
{\langle m_T\rangle}_i\,>\,{\langle m_T\rangle}_{f}, 
\label{fi}
\end{equation}
because of the loss of thermal energy due to work done in longitudinal
expansion. Here the subscript $i$ ($f$) refers to initial (freeze-out)
state.  Assuming ${\langle m_T\rangle}_f\,=\,{\langle m_T\rangle}_{expt}$,
(i.e. all the hadrons are emitted from the freeze-out surface)
and using Eq's.~\ref{eq2} and \ref{fi} we get
\begin{equation}
T_i>{\langle m_T\rangle}_{expt}/2. 
\label{tin}
\end{equation}
A value of $\langle v_r\rangle\,\sim 
0.4$ reproduces the slope of the $m_T$ distribution of pions at SPS energies
reasonably well. For this value of $\langle v_r\rangle$ we get
$T_F\sim 120 $ MeV from Eq.~\ref{blue}, indicating, 
$\langle m_T\rangle=m+T_F+T_F^2/(m+T_F)\,\sim 315$ MeV.  It should be noted here
that the effects of the transverse flow is ``subtracted'' out from
$\langle m_T\rangle$ now. Using Eq.~\ref{tin} we obtain $T_i>158$ MeV.

Photons and dileptons  emitted from the strongly interacting 
matter constitute one of the most efficient probes to measure the
initial temperature of the system~\cite{probe}. 
It was  shown earlier~\cite{interf}
that direct photons (measured by WA98 collaboration~\cite{wa98_photon})
having transverse momentum in the range $1<\, p_T$ (GeV)$<\, 2.5$ 
is expected to 
originate from a thermal source. The inverse slope of this 
spectra is $\sim$ 275 MeV. 
For $\langle v_r\rangle\sim 0.4$ we obtain an average temperature
$T_{av}\sim 178$ MeV by using Eq.~\ref{blue}. A similar value of
$T_{av}$ is obtained from the $p_T$ distribution of $e^+e^-$~\cite{ceres}.
Since photons and dileptons are emitted from all the stages of space
time evolution, $T_{av}$ satisfies the condition, $T_i>T_{av}>T_F$. All
these indicate that a value $T_i\ge 170$ MeV may have been realized 
at SPS energies.

The total multiplicity of hadrons
is related to the initial temperature ($T_{i}$) and thermalization time
($\tau_i$) as:
\begin{equation}
\frac{dN}{dY} 
\sim 4 \frac{\pi^2}{90}\,g_{eff}\, \pi\, R^{2}\, T_{i}^{3}\,\tau_{i}/3.6,
\label{eq3}
\end{equation}
$R$ is the radius of the colliding nuclei and $g_{eff}$ is the effective 
statistical degeneracy. 
Taking $dN/dY \sim 700$ corresponding to the highest SPS
energies and using Eq.~\ref{tin} we get $\tau_i \le 1.2$ fm/c
$\sim 1/\Lambda_{QCD}$ where $\Lambda_{QCD}$ is the QCD scale parameter.

(c) Recently, there have been some attempts to explain the transverse mass
spectra of produced particles without introducing a transverse expansion.
The hadrons gain transverse momentum through successive collisions
with the particles in the system. This can be treated as a 
random walk of the hadrons~\cite{hs} in the medium. The $m_T$ spectra of 
pions at SPS energies can be reproduced by this model, but
it fails to reproduce the slope of kaon and proton distribution
at the same colliding energies. 

(d) As a last consideration, we investigate 
how the ${\langle m_{T}\rangle}$ of produced particles
vary with $\sqrt{s}$ of A-A systems 
in a standard event generator based on the principle
that the A-A collision is a superposition of nucleon-nucleon collisions. This
is done by using the event generator, HIJING~\cite{hijing} which contains
the effect of mini-jets. The results of this calculation do not show the 
variation of ${\langle m_{T}\rangle}$ with $\sqrt{s}$ as has been 
observed experimentally. 

In summary, a detailed analysis of the experimental data of the 
transverse mass spectra of identified hadrons, at AGS, SPS and RHIC energies, 
show a characteristic behaviour  of ${\langle m_{T}\rangle}$ 
as a function of multiplicity similar to that proposed by
Van Hove. The plateau in ${\langle m_{T}\rangle}$ 
corresponding to the multiplicity at SPS energies may hint at the formation 
of a mixed phase of QGP and hadrons in the initial stages. 
The average temperature
obtained from the analysis of photons and dileptons lends support to this.
We have shown by solving the hydrodynamical equations that the
presence of a mixed phase really slows down the growth of $\langle m_T\rangle$
with $dN/dY$ substantially. In the present work all the hadronic degrees of
freedom up to mass 2.5 GeV has been included in the EOS; a fewer hadronic
degrees of freedom will slow down this growth even more. 
Since a lower bound of the initial temperature is obtained here,
the formation of QGP at a higher temperature cannot be ruled out.
However, even if QGP phase is created at SPS, its space time volume
could be too small to affect the dynamics of the system and hence our
conclusions. 
The upward trend in the slope of  ${\langle m_{T}\rangle}$ at RHIC energies
is indicative of a deconfined phase in the initial state.

\vskip -0.4cm
%\acknowledgments
{The authors are thankful to Bikash Sinha for fruitful
discussions. One of us (B.M.) is grateful to the Board of Research
on Nuclear Science and Department of Atomic Energy, 
Government of India for financial support.}

\end{document}